\documentclass{ifacconf}

\usepackage{graphicx}  
\usepackage{natbib}    
\usepackage{enumitem}    

\usepackage{amssymb,amsmath,color,mathtools}

\newtheorem{ex}{Example}

\newcommand{\R}{\mathbb R}

\newcommand{\N}{\mathbb N}

\newcommand{\cA}{\mathcal A}
\newcommand{\cB}{\mathcal B}

\newcommand{\cD}{\mathcal D}

\newcommand{\cF}{\mathcal F}

\newcommand{\cH}{\mathcal H}
\newcommand{\cI}{\mathcal I}

\newcommand{\cK}{\mathcal K}

\newcommand{\cN}{\mathcal N}

\newcommand{\cP}{\mathcal P}

\newcommand{\cS}{\mathcal S}

\newcommand{\cU}{\mathcal U}

\newcommand{\cX}{\mathcal X}
\newcommand{\cY}{\mathcal Y}

\newcommand{\x}{\times}
\newcommand{\st}{\,:\,}
\newcommand{\setto}{\rightrightarrows} 

\newcommand{\inv}{^{-1}}
\newcommand{\uinv}{^{\rm U}}

\renewcommand{\c}{^{\rm c}}

\newcommand{\downcl}[1]{\downarrow\!\!#1}
\newcommand{\upcl}[1]{\uparrow\!\!#1}
\newcommand{\Sigmafb}{\Sigma_{\circlearrowright}}

\definecolor{diffc}{RGB}{200,0,200}

\newcommand{\MB}[1]{{\color{blue}#1}}

\allowdisplaybreaks

\begin{document}
	
\begin{frontmatter}

	\title{On an Abstraction of Lyapunov and Lagrange Stability\thanksref{footnoteinfo}} 
	

	\thanks[footnoteinfo]{\MB{\copyright\ 2025 the
			authors. This work has been accepted to IFAC for publication under a Creative Commons
			Licence CC-BY-NC-ND (NOLCOS 2025).}}
	
	\author[unibo]{Michelangelo Bin} 
	\author[imperial]{David Angeli} 
	
	\address[unibo]{DEI, University of Bologna, Bologna, Italy\\(email:
		\texttt{\emph{michelangelo.bin@unibo.it}})}
		\address[imperial]{EEE, Imperial College London, London, U.K.\\(email: \texttt{\emph{d.angeli@imperial.ac.uk}})}
	
	\begin{abstract} 
		 This paper studies a set-theoretic generalization of Lyapunov and Lagrange stability for abstract systems described by set-valued maps. Lyapunov stability is characterized as the property of inversely mapping filters to filters, Lagrange stability as that of mapping ideals to ideals.
		 These abstract definitions unveil a deep duality between the two stability notions, enable a definition of global stability for abstract systems, and yield an agile generalization of the stability theorems for basic series, parallel, and feedback interconnections, including a small-gain theorem. 
		 Moreover, it is shown that Lagrange stability is abstractly identical to other properties of interest in control theory, such as safety and positivity, whose preservation under interconnections can be thus studied owing to the developed stability results.
	\end{abstract}
	
	
\end{frontmatter}

\section{Introduction}\label{sec.intro}

Lyapunov stability, as originally introduced by A. Lyapunov in 1892  \citep{lyapunov_general_1992}, is a continuity property of the solution map of a differential equation transforming initial conditions in the corresponding trajectories.
Continuity is also a fundamental constituent of input-output notions of stability, as it is either explicitly required  in the definition~\citep{zames_input-output_1966}, or it is implied by stronger notions such as linear gain~\citep{desoer_feedback_1975} or input-to-state stability~\citep{sontag_smooth_1989}. 
Recently, in the context of small-gain analysis, \citep{bin_small-gain_2023} generalized this continuity notion of stability to abstract systems described by set-valued maps between arbitrary topological spaces.
These are systems defined by a triple $(\cD,\cY,\Psi)$ in which $\cD$ and $\cY$ are sets and  $\Psi:\cD\setto \cY$ is a set-valued map; $\cD$ contains the variables considered independent, such as initial conditions, inputs, or parameters, and $\cY$ contains the dependent variables, such as state or output trajectories.
Once $\cD$ and $\cY$ are endowed with some topologies $\tau_\cD$ and $\tau_\cY$, respectively,  \citep{bin_small-gain_2023} defines Lyapunov stability as follows.
\begin{defn}[Lyapunov stability]\label{def.Lyap.topo}
	A subset $D\subset \cD$  is  \emph{(Lyapunov) stable} if, for every $\tau_\cY$-neighborhood $V$ of $\Psi(D)$, there exists a $\tau_\cD$-neighborhood $U$ of $D$, such that $\Psi(U)\subset V$.
\end{defn}
If $D=\{d\}$ is a singleton, this notion reduces to upper-semicontinuity of $\Psi$ at $d$ and, hence, to regular continuity at $d$ if $\Psi$ is single-valued.

It is shown in \citep{bin_small-gain_2023} that, depending on the choice of the spaces $\cD$ and $\cY$ and their topologies, Definition~\ref{def.Lyap.topo} captures the continuity properties behind various common notions of stability, such as Lyapunov stability of sets or motions, input-output stability, asymptotic gain,  and their incremental/integral counterparts. 
Nevertheless, Definition~\ref{def.Lyap.topo} is in general only \emph{local}, as it only involves the neighborhood filters of $D$ and of its image $\Psi(D)$.
In control theory, however, the interest is often toward nonlocal and possibly \emph{global} stability properties, which provide enhanced robustness guarantees. 
The starting point of this paper is the  question of how  such globality requirements can be conceived in the same  setting of \citep{bin_small-gain_2023} described above, thereby obtaining an abstract notion of global stability.

Going back to the classical notion of Lyapunov stability of the origin of an autonomous system, and in particular to its $\delta$-$\epsilon$ formulation,\footnote{The origin of $\dot x=f(x)$ ($x(t)\in\R^n$) is Lyapunov stable if, for every $\epsilon>0$, there exists $\delta(\epsilon)>0$, such that, if $|x(0)|<\delta(\epsilon)$, then $|x(t)|<\epsilon$ for all $t\ge 0$.}
one says that the origin is \emph{globally stable} if $\delta$, as a function of $\epsilon$, can be taken as  defined on the whole $\R_{\ge 0}$ and satisfying $\lim_{\epsilon\to \infty} \delta(\epsilon) = \infty$ \citep{Andriano1997a}. Equivalently,  the origin is globally stable if there exists $\alpha\in\cK_\infty$ such that every solution $x$ of the considered system satisfies
\begin{equation}\label{e.K-stability}
	\forall t\ge 0,\quad |x(t)|\le\alpha(|x(0)|).
\end{equation}
Nevertheless, how the functions $\delta$ and $\alpha$ may be defined in general topological spaces is unclear, and this hinders the direct generalization of the previous definitions.

A third equivalent characterization is that the origin of an autonomous system is \emph{globally stable if and only if it is Lyapunov stable and the system is Lagrange stable} \citep[Thm.~1]{Andriano1997a}.\footnote{A system $\dot x=f(x)$ ($x(t)\in  \R^n$) is Lagrange stable if, for each bounded set $B\subset \R^n$, the positive orbit $\{ x(t)\in \R^n\st \dot x=f(x),\ x(0)\in B \}$ is bounded \citep{Andriano1997a}. See also \citep{loria_stability_2017} for an  historic  perspective.} 
As discussed later in Section~\ref{sec.notions.Lagrange-boundedness}, Lagrange stability can be seen as uniform boundedness; hence, this latter characterization draws a remarkable connection with the input-output definitions of stability: In \citep{zames_input-output_1966} stability is defined as continuity plus boundedness, while both continuity and boundedness are implied by linear gain \citep{desoer_feedback_1975} and input-to-state stability~\citep{sontag_smooth_1989}. 
Moreover, this characterization seems more prone to abstraction, as it only requires a generalization of the notion of Lagrange stability, since Lyapunov stability is already generalized by Definition~\ref{def.Lyap.topo}.

Motivated by the previous observations, this paper explores the generalization of global stability by seeking a set-theoretic abstraction of Lagrange stability along the lines of \citep{bin_small-gain_2023}. In doing so, a step further is taken also with respect to \citep{bin_small-gain_2023} by showing that both Lyapunov and Lagrange stability admit an even more abstract set-theoretic formulation  in terms of the action of the system's map $\Psi$ on suitably-defined ideals and filters (Section~\ref{sec.notions}).
This abstract formulation highlights an elegant duality between the two notions of stability, which is  then leveraged to obtain for Lagrange stability the same stability theorems proved in \citep{bin_small-gain_2023} for Lyapunov stability  of basic series, parallel, and feedback interconnections (Section~\ref{sec.stability}). In particular, these results are stated and proved in the same way with only a slight modification reflecting the duality of the two stability notions.
Finally, it is shown that other properties like \emph{safety} and \emph{positivity}  are abstractly identical to Lagrange stability (with the due selection of the input and output sets). Therefore, the above-mentioned stability results can be  used to study preservation of such properties under interconnections.

 \emph{Notation.}  All norms are denoted by $|\cdot|$. A continuous function $\alpha:\R_{\ge 0}\to \R_{\ge 0}$ is of class-$\cK_\infty$ if it is strictly increasing, $\alpha(0)=0$, and $\lim_{s\to\infty}\alpha(s)=\infty$.
 Let $\cX$ be a set and $X\subset\cX$, then $X\c \coloneq \cX\setminus X$ denotes the complement of $X$.
 If $\cB$ is a family of subsets of  $\cX$, $\downcl{\cB} \coloneq \{ X\subset\cX\st \exists B\in\cB,\, X\subset B \}$ and $\upcl{\cB} \coloneq \{ X\subset\cX\st  \exists B\in\cB,\, B\subset X\}$ denote the \emph{lower} and \emph{upper closure} of $\cB$, respectively.
 Moreover, we let $\cup\cB \coloneq \cup_{B\in\cB} B$.
 Given two families $\cB_1$ and $\cB_2$ of subsets of a set $\cX$, we let $\cB_1\otimes\cB_2\coloneq \{ B_1\x B_2 \subset \cX^2 \st B_1\in\cB_1,\ B_2\in \cB_2\}$.
   $\Psi:\cD\setto\cY$ denotes a set-valued map.  If $D\subset\cD$, we let $\Psi(D)=\cup_{d\in D}\Psi(d)$. If $D=D_1\x D_2$, we let $\Psi(D_1,D_2)=\Psi(D_1\x D_2)$. Given $Y\subset\cY$, we let $\Psi\uinv(Y)\coloneq  \{d\in \cD\st \Psi(d)\subset Y\}$ denote  the \emph{upper inverse} of $\Psi$, and $\Psi\inv(Y)\coloneq  \{d\in \cD\st \Psi(d)\cap Y\ne\varnothing \}$ its \emph{lower inverse}. 

\section{Lyapunov and Lagrange Stability}\label{sec.notions}

This section introduces an abstract notion of Lyapunov and Lagrange stability in terms of set theory. The starting point is the framework of \citep{bin_small-gain_2023} where, as briefly discussed in Section~\ref{sec.intro}, systems are described by triples $(\cD,\cY,\Psi)$, in which $\cD$ and $\cY$ are sets,  and  $\Psi:\cD\setto \cY$ is a set-valued map.

\subsection{Lagrange stability as boundedness}\label{sec.notions.Lagrange-boundedness}
As a first step, this section expresses Lagrange stability in terms of \emph{boundedness} of the map $\Psi$.
Starting with the setting of \citep{Andriano1997a}, let $\cD=\R^n$ be the set of the initial conditions, $\cY$ that of continuous functions $x:\R_{\ge 0}\to \R^n$, and  $\Psi$ be the solution map of some differential equation $\dot x=f(x)$ mapping initial conditions $x_0\in \cD$ to trajectories $x\in\cY$ (for simplicity, we assume that the state trajectories are defined for all times, although this is unnecessary).
Then, the definition of Lagrange stability of \citep{Andriano1997a} can be adapted as follows.
\begin{defn}[Lagrange stability]\label{def.Lagrange-Adriano}
	System $(\cD,\cY,\Psi)$ is \emph{Lagrange stable at $D\subset\cD$} if, for every bounded subset $B$ of $D$, the set $\cup_{x \in \Psi(B)} \cup_{t\ge 0} x(t)$
	is bounded.
\end{defn}
If we define on $\cD$ the Euclidean metric and on $\cY$ the (extended) uniform metric, then Lagrange stability at $D$ can be simply stated as \emph{boundedness of $\Psi$ on $D$}. Clearly, such a definition directly extends to arbitrary sets $\cD$ and $\cY$ on which some (extended) metric can be defined.
However, while Lyapunov stability is a topological property, boundedness is not (see, e.g., \cite[Lem. 3.6]{aliprantis_infinite_2006}).
Thus, Lagrange stability cannot be expressed in terms of the neighborhood filters as, instead, Lyapunov stability was in \citep{bin_small-gain_2023}. 
Rather, a more natural framework to generalize boundedness is that of \emph{bornological spaces} \citep{hogbe-nlend_bornologies_1977}, as bornology abstracts the properties of bounded sets of $\R^n$ like topology does for the open sets. Specifically,
given a set $\cX$, a \emph{bornology on $\cX$} is a collection $\beta$ of subsets of $\cX$ satisfying the following axioms:
\begin{enumerate}[label=\textbf{B\arabic*.},ref=\textbf{B\arabic*}]
	\item $\beta$ covers $\cX$, i.e., $\cup\beta = \cX$.
	\item\label{born.2} If $B_1, B_2 \in \beta$, then $B_1\cup B_2 \in \beta$.
	\item\label{born.3} If $B\in\beta$ and $C\subset B$, then $C\in\beta$.
\end{enumerate} 
The elements of $\beta$ are called \emph{$\beta$-bounded sets}.
With this definition in mind, for a general system $(\cD,\cY,\Psi)$, one can endow $\cD$ and $\cY$ with some bornologies $\beta_\cD$ and $\beta_\cY$ and define Lagrange stability as follows (cf. Definition~\ref{def.Lagrange-Adriano}). 
\begin{defn}[Lagrange stability]\label{def.Lagrange-born}
	System $(\cD,\cY,\Psi)$ is \emph{Lagrange stable at $D\subset\cD$} if $\Psi(U)$ is $\beta_\cY$-bounded for every $\beta_\cD$-bounded $U\subset D$.
\end{defn} 
Definition~\ref{def.Lagrange-born} draws the following correspondence between Lyapunov stability and Lagrange stability
\begin{equation}\label{scheme-lyap-lagr}
	\begin{array}{ccc}
		Lyapunov\ stability &\leftrightarrow & Lagrange\ stability\\[1ex]
		topology & \leftrightarrow & bornology\\[1ex]
		continuity  & \leftrightarrow &  boundedness
	\end{array}
\end{equation}
The ``duality'' \eqref{scheme-lyap-lagr} is a first abstract correspondence between Lyapunov and Lagrange stability. 
The rest of the section is dedicated to a further abstraction of \eqref{scheme-lyap-lagr} in terms of  purely set theory. 
The starting point is to notice that Definitions~\ref{def.Lyap.topo} and \ref{def.Lagrange-born} refer to a given set $D$  although such a dependency is immaterial as the notions are the same for all sets $D$. 
This is a sign that such definitions did not reach the core of the corresponding notions. 
In the spirit of \emph{pointless topology}~\citep{johnstone_point_1983}, the reminder of this section pursues an abstraction of Definitions~\ref{def.Lyap.topo} and~\ref{def.Lagrange-born} that is not affected by such an issue. As a byproduct: \emph{(i)}  duality~\eqref{scheme-lyap-lagr} strengthens and reaches a form that permits to easily port the stability results of \citep{bin_small-gain_2023} to Lagrange stability as briefly commented in Section~\ref{sec.intro}; \emph{(ii)} the notion of Lagrange stability is freed from the meaning of ``boundedness'' thus enabling it to capture other relevant properties such as safety or positivity.

\subsection{Filters and ideals}\label{sec.notions.filters-ideals}
The following definitions are taken from \citep{kuratowski_topology_1966}.
Given a set $\cX$, a \emph{filter} $\cF$ on $\cX$ is a nonempty family of subsets of $\cX$ satisfying the following properties:
\begin{enumerate}[label=\textbf{F\arabic*.},ref=\textbf{F\arabic*}]
	\item\label{filter.1} If $A\in \cF$ and $A\subset B$, then $B\in \cF$.
	\item\label{filter.2} If $A,B\in\cF$, then $A\cap B\in\cF$.
\end{enumerate} 
An \emph{ideal} $\cI$ on $\cX$ is a nonempty family of subsets of $\cX$ satisfying the following properties:
\begin{enumerate}[label=\textbf{I\arabic*.},ref=\textbf{I\arabic*}]
	\item\label{ideal.1}  If $A\in \cI$ and $B\subset A$, then $B\in \cI$.
	\item\label{ideal.2} If $A,B\in\cI$, then $A\cup B\in\cI$.
\end{enumerate} 

Filters and ideals are \emph{dual} in the following sense.
\begin{lem}\label{lem.duality-filter-ideal}
	 $\cF$ is a filter  on a set $\cX$ if and only if $\cI\coloneq \{ A\c\subset\cX\st A \in \cF\}$ is an ideal on $\cX$.
\end{lem} 

While, as proved later in Lemma~\ref{lem.nbd-filter}, filters capture the essence of Lyapunov stability, ideals play a similar role for Lagrange stability. 
Another notion that will be used later in Section~\ref{sec.stability} is that of \emph{filter} and \emph{ideal bases}: A \emph{filter base} (resp. \emph{ideal base}) on a set $\cX$ is a family $\cB$ of subsets of $\cX$ such that $\upcl\cB$ is a filter (resp. $\downcl\cB$ is an ideal). The filter $\upcl\cB$ (resp. ideal $\downcl\cB$) is called the filter (resp. ideal) \emph{generated} by $\cB$. Finally, the next lemma allows us to construct filter and ideal bases on product spaces.
\begin{lem}
	If $\cB_1$ and $\cB_2$ are ideals (resp. filters) on $\cX$, then $\cB_1\otimes \cB_2$ is an ideal (resp. filter) base on $\cX^2$.
\end{lem}
\begin{pf}
	We prove the claim for the case in which   $\cB_1$ and $\cB_2$ are ideals, a similar argument holds for filters.
	We must show that $\downcl{(\cB_1\otimes \cB_2)}$ is an ideal.
	Property~\ref{ideal.1} is obvious. For \ref{ideal.2},  pick $X,Y\in\downcl{(\cB_1\otimes \cB_2)}$, then there exist $B_i^X,B_i^Y\in \cB_i$ ($i=1,2$) such that $X\subset B_1^X\x B_2^X$ and  $Y\subset B_1^Y\x B_2^Y$. Hence, $X\cup Y\subset (B_1^X\x B_2^X)\cup(B_1^Y\x B_2^Y) \subset  (B_1^X\cup B_1^Y)\x (B_2^X\cup B_2^Y)$, which implies $X\cup Y\in \downcl{(\cB_1\otimes \cB_2)}$ as $B_i^X\cup B_i^Y\in\cB_i$ for both $i=1,2$ in view of \ref{ideal.2}.\hfill$\blacksquare$ 
\end{pf}

\subsection{Forward and backward stability}\label{sec.notions.forward-backward}
Let $\cD$ and $\cY$ be sets, $\Psi:\cD\setto\cY$, and let $\cA$ and $\cB$ be families of subsets of $\cD$ and $\cY$, respectively. The concept of \emph{forward} and \emph{backward} \emph{stability} are defined as follows.
\begin{defn}[Forward stability]
	$\Psi$ is \emph{forward $(\cA,\cB)$-stable} if
	\begin{equation*}
		\forall A\in\cA,\quad \Psi(A)\in\cB.
	\end{equation*} 
\end{defn} 
\begin{defn}[Backward stability]
	$\Psi$ is \emph{backward $(\cA,\cB)$-stable} if
	\begin{equation*}
		\forall  B\in\cB,\quad \Psi\uinv(B)\in\cA.
	\end{equation*}  
\end{defn}
Under the partial order defined by non-strict set inclusion, a filter is a meet-semilattice (by \ref{filter.2}) and an ideal is a join-semilattice (by~\ref{ideal.2}). In this case, forward and backward stability are equivalent to the property that the functions 
$\Psi\vert_{\cA}:\cA\to \cB$ and $\Psi\uinv\vert_{\cB}:\cB\to\cA$, defined by $\Psi\vert_{\cA}(A)\coloneq\Psi(A)$ and $\Psi\uinv\vert_{\cB}(B)\coloneq\Psi\uinv(B)$, are well-defined lattice homomorphisms.
\begin{prop}
	Let $\cA$ and $\cB$ be filters (resp. ideals); $\Psi$ is backward (resp. forward) $(\cA,\cB)$-stable if and only if $\Psi\uinv\vert_\cB$ (resp. $\Psi\vert_\cA$) is a semilattice homomorphism.
\end{prop}
\begin{pf}
\emph{(Backward)} 
The ``if'' part directly follows by the fact that $\Psi\uinv\vert_{\cB}$ is defined on all $\cB$.
For the ``only if'' part, notice that
backward $(\cA,\cB)$-stability guarantees that $\Psi\uinv\vert_{\cB}$ is well defined as a map between $\cB$ and $\cA$.
Let $B_1,B_2\in\cB$, then $\Psi\uinv|_{\cB}(B_1\cap B_2) = \Psi\uinv(B_1\cap B_2) = \{ a\in A\st \Psi(a)\subset B_1\cap B_2 \} = \Psi\uinv(B_1)\cap \Psi\uinv(B_2) = \Psi\uinv|_{\cB}(B_1)\cap \Psi\uinv|_{\cB}(B_2)$. Thus, $\Psi\uinv|_{\cB}$ preserves meets.
\emph{(Forward)} Can be established by a symmetric argument.\hfill$\blacksquare$
\end{pf}

Notice that $\Psi$ is backward $(\cA,\cB)$-stable if and only if $\Psi\uinv$ is forward $(\cB,\cA)$-stable.
Hence, backward and forward stability could in principle be reduced to a single definition. Nevertheless, distinguishing the two is more convenient to deal with systems.
If $\Sigma=(\cD,\cY,\Psi)$ is a system, we shall say that $\Sigma$ is forward or backward $(\cA,\cB)$-stable if so is $\Psi$.
Forward and backward stability will be linked to Lyapunov and Lagrange stability in the next two sections.
A \emph{weak} version of such definitions can also be given; see Section~\ref{sec.notions.weak}.

\subsection{Lyapunov stability revisited}

Let $\cX$ be a topological space and let $X\subset\cX$.  
\begin{lem}\label{lem.nbd-filter}
	The set of all neighborhoods of $X$ is a filter.
\end{lem}
\begin{pf}
	We need to verify that the set $\cN(X)$ of all neighborhoods of $X$ satisfies \ref{filter.1} and \ref{filter.2}.  By definition, if $A\in\cN(X)$, it contains an open set $O$ containing $X$. If  $B\subset \cX$ is such that $A\subset B$, then $O\subset B$. Hence, $B\in\cN(X)$. This proves \ref{filter.1}.
	If $A,B\in\cN(X)$, both $A$ and $B$ contain open sets $O_A$ and $O_B$ containing $X$. Then, $O_A\cap O_B$ is open, contains $X$, and is included in $A\cap B$. Hence, $A\cap B\in\cN(X)$. This proves \ref{filter.2}.\hfill$\blacksquare$
\end{pf}
Let $(\cD,\cY,\Psi)$ be a system and let $\cD$ and $\cY$ be endowed with some topologies.
%
\begin{lem}\label{lem.char-stability}
	$(\cD,\cY,\Psi)$ is Lyapunov stable at $D\subset\cD$ (in the sense of Definition~\ref{def.Lyap.topo}) if and only if,  for every neighborhood $V$ of $\Psi(D)$, the set $\Psi\uinv(V)$ is a neighborhood of $D$.
\end{lem}
\begin{pf}
	(\emph{If}) This is obvious since $\Psi(\Psi\uinv(V)) \subset V$ by definition of $\Psi\uinv$.
	(\emph{Only if})
	Let  $V$ be a neighborhood  of $\Psi(D)$.
	As the system is stable at $D$, there exists a  neighborhood $U$ of $D$ such that $\Psi(U)\subset V$.
	By definition of $\Psi\uinv$, $U\subset\Psi\uinv(V)$.
	Hence, by Lemma~\ref{lem.nbd-filter} and \ref{filter.1}, $\Psi\uinv(V)$ is itself a neighborhood of $D$.\hfill$\blacksquare$
\end{pf}

Let $D\subset\cD$, let $\cA$ be the filter of all neighborhoods of $D$, and let $\cB$ be that of all neighborhoods of $\Psi(D)$.
Ultimately, Lemmas~\ref{lem.nbd-filter} and \ref{lem.char-stability} imply the following equivalence.
\begin{thm}\label{thm.Lyap-filters}
	$(\cD,\cY,\Psi)$ is Lyapunov stable at $D$ if and only if $\Psi$ is backward $(\cA,\cB)$-stable.
\end{thm}
Hence, Lyapunov stability is backward $(\cA,\cB)$-stability when $\cA$ and $\cB$ are filters.
The following section   shows that Lagrange stability relates in a similar way to forward stability with respect to ideals.

\subsection{Lagrange stability revisited}
Let $\cX$ be a bornological space and let $X\subset\cX$. 
\begin{lem}\label{lem.bounded-subsets-ideal}
	The set $\cI$ of all bounded subsets of $X$ is an ideal on $\cX$.
\end{lem}
\begin{pf}
	 If $A\in\cI$  and $B\subset A$, then $B$ is bounded (in view of \ref{born.3}) and $B\subset X$; hence, $B\in\cI$, which proves \ref{ideal.1}.
	If $A,B\in\cI$, then $A\cup B$ is bounded by \ref{born.2}, and it is included in $X$; hence, $A\cup B\in\cI$, which proves \ref{ideal.2}.\hfill$\blacksquare$
\end{pf}

Let $(\cD,\cY,\Psi)$ be a system and let $\cD$ and $\cY$ be endowed with bornologies. Pick $D\subset\cD$, and let $\cA$ and $\cB$ denote the collections of bounded subsets of $D$ and $\Psi(D)$, respectively. Then, directly from Definition~\ref{def.Lagrange-born} and Lemma~\ref{lem.bounded-subsets-ideal}, we obtain the following (cf. Theorem~\ref{thm.Lyap-filters}).
\begin{thm}\label{thm.Lagr-ideals}
	$(\cD,\cY,\Psi)$ is Lagrange stable at $D$ if and only if $\Psi$ is forward $(\cA,\cB)$-stable.
\end{thm}
Hence, Lagrange stability is forward $(\cA,\cB)$-stability when $\cA$ and $\cB$ are ideals.

\subsection{Global stability}\label{sec.notions.global-stability}
Theorems~\ref{thm.Lyap-filters} and \ref{thm.Lagr-ideals}  establish the following duality, which is an abstraction of \eqref{scheme-lyap-lagr}: 
\begin{equation*} 
	\begin{array}{ccc}
		Lyapunov\ stability &\ \ \leftrightarrow\ \ & Lagrange\ stability\\[2ex]
		\begin{array}{c}backward\ stability \\
			(wrt\ filters)
		\end{array}  & \ \ \leftrightarrow\ \  & \begin{array}{c}forward\ stability \\
			(wrt\ ideals)
		\end{array} 
	\end{array}
\end{equation*} 
In line with the equivalence 
between global stability and Lyapunov plus Lagrange stability in the case of differential equations, this section proposes to define an abstract notion of global stability in terms of the union of forward and backward stability. Below, $\Sigma=(\cD,\cY,\Psi)$ is a system, $\cF_\cD$ and $\cI_\cD$ are, respectively, a filter and an ideal on $\cD$, and $\cF_\cY$ and $\cI_\cY$ are, respectively, a filter and an ideal on~$\cY$. 
\begin{defn}\label{def.global-stability}
	$\Sigma$ is \emph{$(\cF_\cD,\cI_\cD,\cF_\cY,\cI_\cY)$-globally stable} if it is both backward $(\cF_\cD,\cF_\cY)$-stable and forward $(\cI_\cD,\cI_\cY)$-stable.
\end{defn}
When $\cF_\cD$ and $\cF_\cY$ are the neighborhood filters (in some metric space) of some set $D\subset\cD$ and of its image $\Psi(D)$, and $\cI_\cD$ and $\cI_\cY$ are the ideals of bounded  subsets (with respect to the same metric) of  $\cD$ and $\cY$, then we recover the usual notion of global stability.
In general, however,  $\cD$ and $\cY$ need not be metric spaces and
the involved ideals and filters need not be related by any metric for Definition~\ref{def.global-stability} to make sense.
 Nevertheless, under an additional ``compatibility condition'', which plays the same gluing role of the metric in the canonical case, it is possible to characterize global stability in the sense of Definition~\ref{def.global-stability} in terms of a bound analogous to~\eqref{e.K-stability}.
\begin{defn}\label{def.compatible-IF}
A filter $\cF$ and an ideal $\cI$ on a set $\cX$ are called \emph{compatible} if 
\begin{enumerate}[label=\textbf{H\arabic*.},ref=\textbf{H\arabic*}]
	\item\label{item.compat.id} For every $X\in\cI$, there exists $H\in \cI\cap\cF$, such that $X\subset H$.
	\item\label{item.compat.fi} For every $X\in\cF$, there exists $H\in \cI\cap\cF$, such that $H\subset X$.
\end{enumerate}
\end{defn}
In the metric space case, where $\cI$ is the ideal of bounded subsets of $\cX$ and $\cF$ is the neighborhood filter of some set $X$, the set $\cI\cap\cF$ is the collections of all bounded neighborhoods of $X$. In this case, \ref{item.compat.id} asks that each bounded set is contained in a bounded neighborhood of $X$, and \ref{item.compat.fi}  asks that the bounded neighborhoods of $X$ constitute a neighborhood base of $X$.
Given two collections  $\cA$ and $\cB$ of sets, we let $\cK_\infty(\cA,\cB)$ denote the set of all onto functions $\alpha:\cA\to\cB$  such that, for every $B\in\cB$, there exists $A\in\cA$, such that $\alpha(A)\subset B$.
The following result parallels the characterization \eqref{e.K-stability} of global stability. To ease the notation, we let $\cH_s\coloneq\cI_s\cap \cF_s$ for both $s=\cD,\cY$.
\begin{thm}\label{thm.alpha-stability}
	Suppose that $\cF_\cD$ and $\cI_\cD$ are compatible and $\cF_\cY$ and $\cI_\cY$ are compatible.
	Then, $\Sigma$ is $(\cF_\cD,\cI_\cD,\cF_\cY,\cI_\cY)$-globally stable if and only if there exists $\alpha\in\cK_\infty(\cH_\cD,\cH_\cY)$ such that
	\begin{equation}\label{e.alpha-stab}
		\forall D\in \cH_\cD,\quad \Psi(D)\subset \alpha(D).
	\end{equation}
\end{thm}
\begin{pf}
	\emph{(If)} Pick $D\in\cI_\cD$; by \ref{item.compat.id}, there exists $H\in \cH_\cD$ such that $D\subset H$. Hence, $\Psi(D)\subset\Psi(H)\subset \alpha(H) \in \cI_\cY$, which by \ref{ideal.1} implies $\Psi(D)\in\cI_\cY$. Hence, $\Sigma$ is forward $(\cI_\cD,\cI_\cY)$-stable. Pick now $Y\in\cF_\cY$; by \ref{item.compat.fi}, there exists $V\in\cH_\cY$ such that $V\subset Y$. By definition of $\cK_\infty(\cH_\cD,\cH_\cY)$, there exists $H\in\cH_\cD$ such that $\Psi(H)\subset \alpha(H)\subset V\subset Y$. Hence, $H\subset\Psi\uinv(Y)$, which by \ref{filter.1} implies $\Psi\uinv(Y)\in\cF_\cD$. Thus, $\Sigma$ is backward $(\cF_\cD,\cF_\cY)$-stable.
	
	\emph{(Only if)} Pick $V\in\cH_\cY$; by global stability, $\Psi\uinv(V)\in\cF_\cD$. By \ref{item.compat.fi}, there exists $H\in\cH_\cD$ such that $H\subset\Psi\uinv(V)$; hence, $\Psi(H)\subset V$. For such a set $H$, define $\alpha_0(H)\coloneq V$. In this way, a subset $\cS\subset\cH_\cD$ and a map $\alpha_0:\cS\to\cH_\cY$ are defined so as $\alpha_0$ is onto (as the previous construction holds for each $V\in\cH_\cY$) and satisfies $\Psi(H)\subset\alpha_0(H)$ for every $H\in\cS$. However,  
	$\alpha_0$ is only defined on a possibly strict subset $\cS$ of $\cH_\cD$.
	Pick $H\in\cH_\cD\setminus\cS$. As $H\in\cI_\cD$, then $\Psi(H)\in \cI_\cY$. By \ref{item.compat.id}, there exists $V\in\cH_\cY$, such that $\Psi(H)\subset V$. We set $\alpha_1(H)\coloneq V$, and we finally define 
	\begin{equation*} 
			\forall D\in\cH_\cD,\quad \alpha(D)\coloneq\begin{cases}
				\alpha_0(D)&\text{if}\ D\in\cS\\
				\alpha_1(D)&\text{otherwise}.
			\end{cases} 
	\end{equation*}
	Then, $\alpha:\cH_\cD\to\cH_\cY$ and it satisfies \eqref{e.alpha-stab} by construction. Moreover, $\alpha\in\cK_\infty(\cH_\cD,\cH_\cY)$. Indeed, it is onto as so is $\alpha_0$, and, again by construction of $\alpha_0$, for every $V\in\cH_\cY$, there exists $H\in\cH_\cD$, such that $\alpha(H)=\alpha_0(H)=V\subset V$.  \hfill$\blacksquare$
\end{pf}
We observe that  Theorem~\ref{thm.alpha-stability} implies the following
\begin{equation}\label{e.k-stability}
	\forall d\in \cup\cH_\cD,\quad \Psi(d)\subset \kappa(d)
\end{equation}
in which $\kappa\coloneq\alpha\circ h$ and $h:\cup\cH_\cD\to \cH_\cD$ is any function such that, for every $d\in\cup\cH_\cD$, $d\in h(d)$. Existence of a function $h$ with such a property follows since $d\in\cup\cH_\cD$ if and only if there exists $D\in\cH_\cD$ such that $d\in D$.
Notice that Inclusion~\eqref{e.k-stability} is analogous to~\eqref{e.K-stability}.


\subsection{Forward stability beyond Lagrange stability}\label{sec.notions.beyond-Lagrange}

In a bornological space $\cX$, boundedness is a property of the subsets of $\cX$ that is \emph{upward directed}, i.e., each two bounded sets are contained in a common larger bounded set.
In general, properties of this kind can be interpreted as ``uniform properties'' of the points of $\cX$.
For instance, the elements of a bounded set share the same bound, which is therefore a uniform bound for all the elements of the set, and a common bound exists for the elements of each pair of bounded sets.
Formally, let $\cX$ be a set; a \emph{uniform property} $\cP$ on $\cX$ is a collection $\cP\subset 2^\cX$ of subset of $\cX$ that is upward directed, i.e., for every $A,B\in\cP$, there exists $C\in\cP$, such that $A\subset C$ and $B\subset C$. 
Ideals provide a set-theoretic characterization of uniform properties in the following terms. 

\begin{lem}\label{lem.boundedness-P}
Every uniform property is an ideal base. 
\end{lem}
\begin{pf}
	Let $\cX$ be a set and $\cP$ a uniform property on it. Let $A\in\downcl{\cP}$; then there exists $P\in\cP$ such that $A\subset P$. If $B\subset A$, then, $B\subset A\subset P$, so as $B\in \downcl{\cP}$. This proves that $\downcl{\cP}$ satisfies \ref{ideal.1}.
	Let $A,B\in\downcl{\cP}$ and $S_a,S_b\in\cP$ be such that $A\subset S_a$ and $B\subset S_b$. As $\cP$ is upward directed, there exists $C\in\cP$ such that $S_a\cup S_b\subset C$. This implies $A\cup B\subset C$ and, thus, $A\cup B\in\downcl{\cP}$. Hence, $\downcl{\cP}$ satisfies~\ref{ideal.2}.\hfill$\blacksquare$
\end{pf} 
By Lemma~\ref{lem.boundedness-P}, 
the collection $\downcl{\!\cP}$
is an ideal to which we refer as the ideal \emph{generated by $\cP$}.
As previously mentioned, if $\cX$ is a bornological space and $\beta$ is a bornology on $\cX$, boundedness is described by the property $\cP \coloneq  \beta$, which defines the ideal of bounded subsets.
Outside boundedness, there are many other uniform properties of interest in control theory that can be defined on $\cX$, and  the notion of forward stability developed in the previous sections as well as the stability results presented in the next sections directly apply to all of them. 
By way of example, the following two examples discuss the cases of positivity and safety.
\begin{ex}[Positive systems]
	Let $\cX$ be a \emph{Riesz space} with order $\ge$; 
	  the \emph{positivity property} is defined as
	\begin{equation*}
		\cP \coloneq  \{ X\subset\cX\st \forall x\in X,\, x\ge 0\}.
	\end{equation*}
	Consider the differential equation
	\begin{equation}\label{s.x}
		\dot x = f(x)
	\end{equation}
	with $x(t)\in \cD\coloneq\R^n$ (with the usual componentwise order), and let $\cY$ be the space of continuous functions $\R_{\ge 0}\to\R^n$ endowed with the pointwise and componentwise order (again, for simplicity, we assume that all solutions of \eqref{s.x} from every initial condition are forward complete).
	Let $\cA$ and $\cB$ be the ideals generated by the positivity property on $\cD$ and $\cY$, respectively. Then, system \eqref{s.x} is positive (in the sense of \citep{luenberger_introduction_1979,de_leenheer_stabilization_2001,angeli_monotone_2003}) if and only if it is forward $(\cA,\cB)$-stable.
\end{ex}  
\begin{ex}[Safety]
	Let $\cX$ be a set, and let $\cP\subset2^\cX$ denote a set of \emph{safe} regions within $\cX$.
	We shall assume that, if $S_1,S_2\in\cP$ are safe, then so is their union $S_1\cup S_2$. Hence, $\cP$ is upward directed and therefore it is a uniform property which we call the \emph{safety property}.
	Consider again system \eqref{s.x} and assume that some safety properties $\cP_\cD$ and $\cP_\cY$ have been defined on $\cD$ and $\cY$, respectively.  Let $\cA$ and $\cB$ be the ideals generated by $\cP_\cD$ and $\cP_\cY$, respectively. Then,  system \eqref{s.x} is called \emph{safe} if and only if it is forward $(\cA,\cB)$-stable, namely, if it maps safe regions of initial conditions into safe sets of trajectories.
\end{ex}

\subsection{Weak stability}\label{sec.notions.weak}
Forward/backward stability, as defined in Section~\ref{sec.notions.forward-backward}, relate to \emph{strong} stability requirements, in the sense that \emph{all} images/preimages of $\Psi$ are required to satisfy the corresponding stability notion. Examples of weak stability, not captured by forward and strong stability, are given below.
\begin{ex}[Weak Lyapunov stability]
	Consider the difference inclusion:
	\begin{equation}\label{s.ex.y}
		y^{t+1}  \begin{cases}
			\in \{0,p\} & \text{if}\ y^t \in[0,p/2]\\
			=\frac 1 2 y^t & \text{otherwise}
		\end{cases}
	\end{equation}
	where $p\ge 0$ is a parameter. 
	Let $\cD\coloneq \R_{\ge 0}$ (with the relative Euclidean topology) be the state space, $\cY$ the space of bounded functions $\N\to \R_{\ge 0}$ (with the uniform norm), and $\Psi$ be the solution map of \eqref{s.ex.y}. 
	If $p=0$, the set $D=\{0\}$ is Lyapunov stable; if $p>0$, it is not.
	However, $D$ is always \emph{weakly Lyapunov stable} in the sense that, for every $\epsilon>0$, there exists $\delta>0$, such that, for every initial condition satisfying $|y^0|<\delta$, at least one corresponding solution always satisfies $|y^t|<\epsilon$ for all $t\ge 0$.  
\end{ex}
\begin{ex}[Weak Lagrange stability]
	Consider the differential inclusion
	\begin{equation}\label{s.ex.y2}
		\dot y(t) \in [-y(t),y(t)].
	\end{equation}
	Let $\cD\coloneq \R_{\ge 0}$ be the state space, $\cY$ the space of continuous functions $\R_{\ge 0}\to \R$, and $\Psi$ be the solution map of \eqref{s.ex.y2}. 
	From every initial condition $y(0)>0$,  \eqref{s.ex.y2} has a at least one bounded solution  and at least one unbounded one. The bounded solutions originating from a bounded subset of initial conditions can be uniformly bounded.  Hence, we say that $D=\cD$ is weakly Lagrange stable.
\end{ex}
Weak stability can be defined in the language of Section~\ref{sec.notions.forward-backward} as follows.
Below, $\cD$ and $\cY$ are sets, $\Psi:\cD\setto\cY$, and   $\cA$ and $\cB$ are families of subsets of $\cD$ and $\cY$, respectively.  
\begin{defn}[Weak forward stability]
	$\Psi$ is \emph{weakly forward $(\cA,\cB)$-stable} if
	\begin{equation*}
		\forall A\in\cA,\ \exists B\in\cB,\ \Psi(A)\cap B\ne \varnothing.
	\end{equation*}
\end{defn}
 
\begin{defn}[Weak backward stability]
	$\Psi$ is \emph{weakly backward $(\cA,\cB)$-stable} if  
	\begin{equation*}
		\forall B\in\cB,\ \Psi\inv(B)\in\cA,
	\end{equation*} 
\end{defn} 
denotes $\Psi\inv$ is the lower  inverse of $\Psi$. 

\section{Stability of Interconnections}\label{sec.stability}
This section presents some basic forward and backward stability results for series, parallel, and feedback interconnections; the latter provide a generalization of the small-gain theory developed in \citep{bin_small-gain_2023} for Lyapunov stability.
Whenever possible, we   develop the results for generic families of sets or for filter/ideal bases. Indeed: \emph{(i)}   filters (resp. ideals) are filter bases (resp. ideal bases); \emph{(ii)} backward (resp. forward) stability with respect to filter bases (resp. ideal bases) implies backward (resp. forward) stability with respect the filter (resp. ideal)  generated  by the base.
Item \emph{(i)} holds since if $\cB$ is a filter (resp. an ideal), then $\upcl{\cB}=\cB$ (resp. $\downcl{\cB}=\cB$) in view of \ref{filter.1} and \ref{ideal.1}. Item \emph{(ii)} follows from the next lemma.
\begin{lem}
	Let $\cA$ and $\cB$ be ideal (resp. filter) bases; if $\Psi$ is forward (resp. backward) $(\cA,\cB)$-stable, it is also forward $(\downcl{\cA},\downcl{\cB})$-stable (resp. backward $(\upcl{\cA},\upcl{\cB})$-stable).
\end{lem}
\begin{pf}
	\emph{(Forward)}	For every $A\in\downcl{\cA}$, there exists $\bar A\in\cA$ such that $A \subset \bar A$. Forward stability implies $\Psi(A)\subset  \Psi(\bar A)\in\cB$. Therefore, $\Psi(A)\in\downcl{\cB}$.
	\emph{(Backward)}	If $B\in\upcl{\cB}$, there exists $\bar B\in\cB$ such that $\bar B \subset B$. Backward stability implies $\Psi\uinv(B)\supset \Psi\uinv(\bar B) \in\cA$. Hence, $\Psi\uinv(B)\in\upcl{\cA}$.\hfill $\blacksquare$
\end{pf}

\subsection{Series interconnections}
Consider two systems $\Sigma_1=(\cD_1,\cY_1,\Psi_1)$ and $\Sigma_2=(\cD_2,\cY_2,\Psi_2)$ with $\cY_1\subset\cD_2$.
The \emph{series interconnection of $\Sigma_1$ and $\Sigma_2$} is the system $\Sigma_1\Sigma_2=(\cD,\cY,\Psi)$ with $\cD\coloneq\cD_1$, $\cY\coloneq\cY_2$, and $\Psi(d)\coloneq \Psi_2(\Psi_1(d))$ for all $d\in\cD$. 
The series interconnections of Lyapunov stable systems is Lyapunov stable \citep[Prop. 1]{bin_small-gain_2023}. Hereafter, such a result is extended to general backward and forward stability. Below, for $i=1,2$, $\cA_i$ and $\cB_i$ are generic families of subsets of $\cD_i$ and $\cY_i$, respectively. 
\begin{thm}\label{thm.series.back}
	Assume that $\Sigma_1$ is backward (resp. forward) $(\cA_1,\cB_1)$-stable, $\Sigma_2$ is backward (resp. forward) $(\cA_2,\cB_2)$-stable, and $\cA_2\subset \cB_1$ (resp. $\cB_1\subset\cA_2$). Then, $\Sigma_1\Sigma_2$ is backward (resp. forward) $(\cA_1,\cB_2)$-stable.
\end{thm}
\begin{pf}
	\emph{(Backward)} Pick $B\in\cB_2$. Then $\Psi_2\uinv(B)\in \cA_2 \subset \cB_1$. Hence, $\Psi_1\uinv(\Psi_2\uinv(B))\in \cA_1$.
	The result then follows by the fact that $(\Psi_2\circ\Psi_1)(\cdot)\uinv =\Psi_1\uinv(\Psi_2\uinv(\cdot))$; indeed $x\in (\Psi_2\circ\Psi_1)\uinv(B) \iff \Psi_2(\Psi_1(x))\subset B \iff \Psi_1(x)\subset\Psi_2\uinv(B)\iff x\in\Psi_1\uinv(\Psi_2\uinv(B))$.
	\emph{(Forward)} For every $A\in\cA_1$, one has $\Psi_1(A)\in\cB_1\subset\cA_2$. Hence, $\Psi_2(\Psi_1(A))\in \cB_2$.\hfill$\blacksquare$
\end{pf}
Theorem~\ref{thm.series.back} connects to Lyapunov stability of the series (hence, to \citep[Prop. 1]{bin_small-gain_2023}) whenever $\cA_1$ and $\cB_1$ are the neighborhood filters of some set $D\subset\cD_1$ and of $\Psi_1(D)$, respectively, and $\cA_2$ and $\cB_2$ are the neighborhood filters of $\Psi_1(D)$ and $\Psi_2(\Psi_1(D))$. The condition $\cA_2\subset\cB_1$ means that $\cB_1$ is \emph{finer} than $\cA_2$; hence, the topologies in question are compatible. Under such conditions, the backward part of Theorem~\ref{thm.series.back} reads as: if $\Sigma_1$ is Lyapunov stable at $D$ and $\Sigma_2$ is Lyapunov stable at $\Psi_1(D)$, then $\Sigma_1\Sigma_2$ is Lyapunov stable at $D$. 
Instead, when $\cA_1$, $\cB_1$, $\cA_2$, $\cB_2$ are ideals, and the compatibility condition $\cB_1\subset \cA_2$ holds,  Theorem~\ref{thm.series.back} states that the series interconnection of Lagrange stable systems is Lagrange stable. 
According to Section~\ref{sec.notions.beyond-Lagrange}, the latter result also applies to safety, positivity, and other properties.

\subsection{Parallel Interconnections}

 Consider two systems $\Sigma_1=(\cD_1,\cY_1,\Psi_1)$ and $\Sigma_2=(\cD_2,\cY_2,\Psi_2)$ with $\cD_1=\cD_2$.
The \emph{parallel interconnection of $\Sigma_1$ and $\Sigma_2$} is the system $\Sigma_1\x\Sigma_2=(\cD,\cY,\Psi)$ with $\cD\coloneq\cD_1=\cD_2$, $\cY\coloneq\cY_1\x\cY_2$, and $\Psi(d)\coloneq \Psi_1(d)\x \Psi_2(d)$ for all $d\in\cD$.
The following result states that parallel interconnections of forward/backward stable systems are stable.  Below, for $i=1,2$, $\cA_i$ and $\cB_i$ denote families of subsets of $\cD_i$ and $\cY_i$, respectively. 
\begin{thm}\label{thm.parallel}
	Let $\cA_i$  be an ideal (resp. filter) base.
	Assume that, for $i=1,2$, $\Sigma_i$ is forward (resp. backward) $(\cA_i,\cB_i)$-stable, and $\cA_1\subset\cA_2$. Then, $\Sigma_1\x \Sigma_2$ is forward $(\cA_1, \downcl{(\cB_1\otimes \cB_2)})$-stable (resp. backward $(\upcl{\cA_2}, \cB_1\otimes \cB_2)$-stable).
\end{thm}
\begin{pf}
\emph{(Forward)} Since $\cA_1\subset\cA_2$, and in view of forward stability of $\Sigma_1$ and $\Sigma_2$,   for every $A\in \cA_1$  we have $\Psi(A) = \cup_{d\in  A} (\Psi_1(d)\x\Psi_2(d)) \subset \Psi_1(A)\x \Psi_2(A)   \in \cB_1\otimes\cB_2$, which implies $\Psi(A)\in \downcl{(\cB_1\otimes\cB_2)}$.
\emph{(Backward)}  Pick $B_1\x B_2\in  \cB_1\otimes \cB_2$; then, $\Psi\uinv(B_1\x B_2)   = \{ d\in\cD\st \Psi_1(d)\x\Psi_2(d)\subset B_1\x B_2\} = \{ d\in\cD\st \Psi_1(d) \subset B_1 \} \cap \{ d\in\cD\st \Psi_2(d) \subset B_2 \}= \Psi_1\uinv(B_1)\cap\Psi_2\uinv(B_2)$.
As  $\Sigma_i$ is backward $(\cA_i,\cB_i)$-stable, and $\cA_1\subset\cA_2$, then  $\Psi_i\uinv(B_i)\in \cA_2$ for both $i=1,2$.
Since $\cA_2$ is a filter base, $\upcl{\!\cA_2}$ is a filter, and \ref{filter.2} implies $\Psi\uinv(B_1\x B_2) =\Psi_1\uinv(B_1)\cap \Psi_2(B_2)\in \upcl{\cA_2}$. \hfill$\blacksquare$
\end{pf}

Theorem~\ref{thm.parallel} parallels Theorem~\ref{thm.series.back} for parallel interconnections. However, while the abstract statements have the same form, there is a relevant difference for what concerns the implications for Lyapunov stability: As noticed in \citep{bin_small-gain_2023}, while Lyapunov stability of a series holds for any set $D$,  Lyapunov stability of a parallel interconnection of the considered kind is only generically true for singletons (i.e., $D=\{d\}$ for some $d\in \cD$). A counterexample is proposed below. The reason is that, if $D=\{d\}$ is a singleton, then $\Psi(d)=\Psi_1(d)\x \Psi_2(d)$ is a rectangle; hence, $\cB_1\otimes\cB_2$ generates the neighborhood filter of $\Psi(d)$. However, if $D$ is not a singleton, $\Psi(d) = \cup_{d\in D} \Psi_1(d)\x \Psi_2(d)$ may not be a rectangle; in this case, $\cB_1\otimes\cB_2$ does not in general generate the neighborhood filter of $\Psi(d)$. 
\begin{ex}
Let $\Sigma_i=(\cD,\cY,\Psi_i)$ where $\cD=\cY=\R_{\ge 0}$ with relative Euclidean topology, $\Psi_1(d)\coloneq d$, and $\Psi_2(d)\coloneq d$ if $d\in[0,1]$ and $\Psi_2(d)\coloneq  0$ otherwise. 
Both systems are Lyapunov stable at $D=[0,1]$.
However, the parallel interconnection is not. To see this, notice that $\Psi(D) = \cup_{d\in[0,1]} \Psi_1(d)\x \Psi_2(d) = \{ (y_1,y_2)\in [0,1]^2\st y_1=y_2\}$ and that $V\coloneq \{ v\in(\R_{\ge 0})^2\st \inf_{y\in\Psi(D)}|v-y| < 1/2\}$ is a neighborhood of $\Psi(D)$ that satisfies $\Psi\uinv(V) = \{ d\in\R_{\ge 0}\st  \inf_{y\in[0,1]} |(y-\Psi_1(d),y-\Psi_2(d))|<1/2\} \subset [0,1]$. 
The last inclusion follows by contradiction: 
Let $d\in\Psi\uinv(V)$ satisfy $d>1$. Then, for every $\epsilon>0$, there exists $y\in[0,1]$ such that  $|(y-\Psi_1(d),y-\Psi_2(d))|=|(y-d,y)|<1/2+\epsilon$. This, in turn, implies $y<1/2+\epsilon$ and $|y-d|=d-y<1/2+\epsilon$; hence $d<1+2\epsilon$. For $\epsilon$ small enough, we thus obtain a contradiction. 
Since $[0,1]=D$ does not belong to the neighborhood filter of itself, then Lyapunov stability does not hold. 
\end{ex}

\subsection{Feedback interconnections}
Consider two systems
$\Sigma_i =  (\cD_i,\cY_i,\Psi_i)$ in which    $\cD_i = \cY_j\x \cU_i$ for some set $\cU_i$ and for both $(i,j)=(1,2),(2,1)$.
The feedback interconnection 
 of $\Sigma_1$ and $\Sigma_2$ is the system  
$\Sigmafb=(\cD,\cY,\Psi)$ with $\cD\coloneq \cU_1\x \cU_2$, $\cY\coloneq \cY_1\x \cY_2$,   and, for each $d=(d_1,d_2)\in\cD$,
\begin{equation*}
	\begin{aligned}
		\Psi(d)\coloneq \big\{ (y_1,y_2)\in\cY\st  y_1&\in\Psi_1(y_2,d_1),\\ y_2&\in\Psi_2(y_1,d_2) \, \big\}.
	\end{aligned} 
\end{equation*}
For every $d\in\cD$, define the projections
\begin{align*}
	\Upsilon_1(d)&\coloneq \{ y_1\in\cY_1\st \exists y_2\in\cY_2,\ (y_1,y_2)\in\Psi(d) \} \\
	\Upsilon_2(d)&\coloneq \{ y_2\in\cY_2\st \exists y_1\in\cY_1,\ (y_1,y_2)\in\Psi(d) \}.
\end{align*}
For $(i,j)\in\{(1,2),(2,1)\}$, define the maps $\Gamma_{ij}:\cY_i\x \cD \setto \cY_i$ as
\begin{equation*}
	\Gamma_{ij}(y_i,d) \coloneq  \Psi_i(\Psi_j(y_i,d_j),d_i),
\end{equation*}
and, for all $n\in\N$, define $\Gamma_{ij}^n$  according to the recursion
\begin{align*}
	\Gamma_{ij}^1(y_i,d) &\coloneq  \Gamma_{ij}(y_i,d)\\
	\Gamma_{ij}^{n+1}(y_i,d)&\coloneq  \Gamma_{ij}(	\Gamma_{ij}^{n}(y_i,d),d), & n&\ge1.
\end{align*}
The following lemma follows from \citep{bin_small-gain_2023}.
\begin{lem}\label{lem.fbi}
	The following properties hold:
	\begin{enumerate}[label=\textbf{C\arabic*.}, ref=\textbf{C\arabic*}]
		\item\label{fbi.1} $\forall d\in\cD$, $\Psi(d)\subset \Upsilon_1(d)\x \Upsilon_2(d)$.
		\item\label{fbi.2} For every $(i,j)\in\{(1,2),(2,1)\}$, $D\subset\cD$, $y_i\in\Upsilon_i(D)$, and $n\in\N_{\ge 1}$, it holds that $y_i\in\Gamma_{ij}^n(y_i,D)$.
	\end{enumerate}
\end{lem}
\begin{pf}
	\ref{fbi.1} is obvious, while $\ref{fbi.2}$ is \citep[Lem. 4]{bin_small-gain_2023}.\hfill$\blacksquare$
\end{pf}
 
In \citep{bin_small-gain_2023}, a sufficient ``small-gain'' condition for Lyapunov stability of feedback interconnections  with respect to arbitrary topologies on $\cD$ and $\cY$ is given in terms of a  contraction property of the maps $\Gamma_{12}^n$ and $\Gamma_{21}^n$ for large enough $n$ (see \citep[Def.~3 and Thm. 1]{bin_small-gain_2023}). Such a property is shown to be implied by canonical small-gain conditions in the context of input-to-state or input-output stable systems.
Here, we extend this result to handle the more abstract  backward stability property, and we also derive a symmetric result for forward stability.
Below, we consider a feedback interconnection $\Sigmafb=(\cD,\cY,\Psi)$ defined as above, and $\cA$ and $\cB$  denote families of subsets of $\cD$ and $\cY$, respectively. 

\begin{defn}[Small-gain property]\label{def.small-gain}
$\Sigmafb$ is said to satisfy the \emph{backward} (resp. \emph{forward}) \emph{small-gain property with respect to $(\cA,\cB)$} if, for each $B\in\cB$ (resp. $A\in\cA$), there exists $A\in\cA$ (resp. $B\in\cB$), such that
\begin{equation}\label{back-sgp}
	\begin{aligned}
		&\forall (y_1,y_2)\in B\c,\ \exists n_1,n_2\in\N,\\
		&\qquad \text{s.t.} \ \Gamma_{12}^{n_1}(y_1,A)\x \Gamma_{21}^{n_2}(y_2,A)\subset B.
	\end{aligned} 
\end{equation}
\end{defn}
Along the lines of \cite[Thm. 1]{bin_small-gain_2023}, we can then prove the following \emph{small-gain theorem}, showing that the small-gain property of Definition~\ref{def.small-gain} is sufficient for stability.
\begin{thm}\label{thm.sg-back}
	If $\Sigmafb$ satisfies the backward (resp. forward) small-gain property with respect to $(\cA,\cB)$, then it is backward  $(\upcl\cA,\cB)$-stable (resp. forward $(\cA,\downcl\cB)$-stable).
\end{thm}
\begin{pf}
	\emph{(Backward)} The proof mimics that of \citep[Thm.~1]{bin_small-gain_2023}.
	Pick $B\in\cB$ arbitrarily, and let $A\in\cA$ be such that  \eqref{back-sgp} holds.
	Item \ref{fbi.1} of Lemma \ref{lem.fbi} implies
	\begin{align}
		&\Psi(A)\subset \bigcup_{a\in A} \Upsilon_1(a)\x \Upsilon_2(a) \subset \Upsilon_1(A)\x \Upsilon_2(A)\nonumber
		\\
		&\ = \Big((\Upsilon_1(A)\x \Upsilon_2(A))\setminus B\Big) \cup \Big( (\Upsilon_1(A)\x \Upsilon_2(A))\cap B \Big)\nonumber
		\\&\ \subset \Big((\Upsilon_1(A)\x \Upsilon_2(A))\setminus B\Big) \cup B .\label{e.sgt.1}
	\end{align}
	For every $(y_1,y_2)\in (\Upsilon_1(A)\x \Upsilon_2(A))\setminus B$, Item \ref{fbi.2} of Lemma \ref{lem.fbi} implies 
	\begin{equation}\label{e.yi_in_Gammaij}
		y_i \in \Gamma_{ij}^{n_i}(y_i,A), \quad \forall n_i\ge 1.
	\end{equation}
	For each $(y_1,y_2)\in (\Upsilon_1(A)\x \Upsilon_2(A))\setminus B$, let $n_1$ and $n_2$ be such that \eqref{back-sgp} holds. Then, \eqref{back-sgp} and \eqref{e.yi_in_Gammaij} imply
	\begin{align*}
		&(\Upsilon_1(A)\x \Upsilon_2(A))\setminus B = \bigcup_{y \in (\Upsilon_1(A)\x \Upsilon_2(A))\setminus B} \{y_1\}\x \{y_2\}\\
		&\ \subset \bigcup_{y \in (\Upsilon_1(A)\x \Upsilon_2(A))\setminus B} \Gamma_{12}^{n_1}(y_1,A)\x \Gamma_{21}^{n_2}(y_2,A) \ \subset B.
	\end{align*}
	From \eqref{e.sgt.1}, we then obtain $\Psi(A)\subset B$, which implies $A\subset \Psi\uinv(B)$.
	As  $A\in\cA$,  then   $\Psi\uinv(B)\in\upcl\cA$.
	
	\emph{(Forward)} 	Pick $A\in\cA$ arbitrarily, and let $B\in\cB$ be such that  \eqref{back-sgp} holds. Proceeding as above, we obtain again that $\Psi(A)\subset B$, which implies $\Psi(A)\in\downcl{\cB}$.\hfill$\blacksquare$
\end{pf}

We remark that, when $\cA$ and $\cB$ are neighborhood filters, Theorem~\ref{thm.sg-back} gives a small-gain theorem for Lyapunov stability; when $\cA$ and $\cB$ are ideals of bounded sets, it gives a small-gain theorem for Lagrange stability. However, $\cA$ and $\cB$ are not required to have such properties; thus, by way of example, if $\cA$ is a neighborhood filter and $\cB$ a safety ideal, Theorem~\ref{thm.sg-back} gives a robust safety result.

\section{Conclusion}
This paper explored a set-theoretic abstraction of the notion of global stability for systems defined as set-valued maps between sets. 
The proposed notion is the union of forward and backward stability, the former generalizing Lyapunov stability and the latter Lagrange stability. 
The pursued abstraction reveals a deep duality between the two stability notions, which is exploited to prove some generalized stability theorems for basic interconnections.
Future research will be devoted to employ the developed theory to analyze and characterize \emph{robust} stability in the same spirit of  \citep{georgiou_robustness_1997}, which finds application in the  study of learning-based approximation of system models and controllers.

\bibliography{biblio}

\end{document}